# Anomalous CDW ground state in $Cu_2Se$: a wave-like fluctuation of *dc* I-V curve near 50 K

Mengliang Yao[1‡], Weishu Liu[2‡], Xiang Chen[1], Zhensong Ren[1], Stephen Wilson[3], Zhifeng Ren[2*], Cyril P. Opeil[1*]

[1] Department of Physics, Boston College, Chestnut Hill, MA 02467

[2] Department of Physics and TcSUH, University of Houston, TX 77204

[3] Department of Materials, University of California, Santa Barbara, CA 93106

[‡] Equal contributor

[*] Corresponding author: zren@uh.edu, opeil@bc.edu

**Abstract**

A charge density wave (CDW) ground state is observed in polycrystalline $Cu_2Se$ below 125 K, which corresponds to an energy gap of 40.9 meV and an electron-phonon coupling constant of 0.6. Due to the polycrystalline structure, the Peierls transition process has been expanded to a wide temperature range from 90 to 160 K. The Hall carrier concentration shows a continuous decrease from $2.1 \times 10^{20}$ to $1.6 \times 10^{20}$ $cm^{-3}$ in the temperature range from 160 K to 80 K, while almost unchanged above 160 K and below 90 K. After entering the CDW ground state, a wave-like fluctuation was observed in the I-V curve near 50 K, which exhibits as a periodic negative differential resistivity in an applied electric field due to the current. We also investigated the doping effect of Zn, Ni, and Te on the CDW ground state. Both Zn and Ni doped $Cu_2Se$ show a CDW character with increased energy gap and electron-phonon coupling constant, but no notable Peierls transition was observed in Te doped $Cu_2Se$. Similar wave-like I-V curve was also seen in $Cu_{1.98}Zn_{0.02}Se$ near 40 K. The regular fluctuation in *dc* I-V curve was not magnetic field sensitive, but temperature and sample size sensitive.

Charge density wave (CDW) is a periodic modulation of the electronic charge density with an accompanying lattice structure distortion, resulting from the formation of electron-hole pair at the Fermi surface due to the instability of the lattice structure, which is usually induced by a perfect nesting on the Fermi surface. The pair formation is very similar to the Cooper pair formation in superconductors, and the BCS description and estimation are suitable here.[1-3] The appearance of CDW fluctuation is considered as a continuous transition, *i.e.*, Peierls transition, due to the re-arrangement of electronic wave functions driven by the strong phonon-electron interactions at low temperature. The CDW is usually observed in single crystalline materials with layered or chain sub-crystalline structures, such as $NbSe_3$ [4-6], $TaS_3$ [7], $K_{0.3}MoO_3$ [8], $(TaSe_4)_2I$ [9], $KNi_2S_2$ [10] and $La_{1.9}Sr_{0.1}CuO_4$ [11]. Due to the appearance of CDW, abnormal temperature dependent electrical resistivity and also nonlinear I-V curve beyond the threshold electric field were observed [4-9]. Additionally, the coupling effect between electron and phonon was also confirmed by the magnetic susceptibility [12] and heat capacity measurements [13]. However, the ground state with charge density wave is quite weak, which could be broken or depinned by electric field [5-9], magnetic field [14], and pressure [4]. The CDW fluctuation is, therefore, experimentally observed only in high pure single crystalline thin samples such as films or whiskers.

The CDW fluctuation could be smeared out as the materials going from 2D-film to 3D-bulk [2], and also as the materials going from single crystals to polycrystals.[15] Here, we reported a new CDW ground state was observed in $Cu_2Se$ in a polycrystalline bulk near 125 K. The coupling strength was interpreted by the amplification of the resistivity hump near the transition temperature. The Hall effect measurements suggest the Peierls transition starting near 160 K and ending near 90 K. After $Cu_2Se$ enters a new CDW ground state, a negative differential resistivity is observed in the *dc* I-V curve within the critical electric field near 50 K for the first time.

The bulk polycrystalline $Cu_2Se$ samples were synthesized by mechanical ball milling (BM) and hot pressing (HP) at 700 °C. The synthesis technique is similar to our previous work [16]. The typical grain size of as-fabricated sample is on the order of 1-2 μm. LR-700 AC Bridge from Linear Research Inc. is used to perform the electrical resistivity and Hall coefficient measurements from 5 to 300 K. The *dc* I-V curve of the sample is measured by the standard 4-probe method with Keithley 224 current source, and Agilent 34410A 6½ Digit multimeter or Keithley 2182A nanovoltmeter. In all temperature dependent experiments the temperature is controlled and monitored by PPMS with a sweeping speed of 0.5 K/min.

Fig. 1(a) shows the powders XRD pattern of the polycrystalline $Cu_2Se$ made by ball milling and hot pressing. The XRD measurement was conducted on a Brucker D2 PHASER system at room temperature with a scanning speed of 10 seconds per/step (step size = 0.014°). Rietveld refinement was done in Fullprof suite by using monoclinic structure (space group C2/c, No. 15) as the starting structure, which was recently proposed by Gulay [17]. The monoclinic α-$Cu_2Se$ shows a lamella sub-lattice structure along ab-plane, which is similar to its high temperature cubic β-$Cu_2Se$ [16]. The interesting feature of each lamella is the hexagonal-ring chain made by $Cu_3Se_3$, which is outlined in stick mode crystalline structure, as shown in the Fig1. (b). However, the direction of connected $Cu_3Se_3$ chain between two lamellas is different. The one-dimensional hexagonal ring chain could generate the electronic density instability associated with the electron-phonon coupling effect. Fig. 1(c) shows the temperature dependent electrical resistivity of a typical polycrystalline $Cu_2Se$. Reversible non-linear temperature dependent electrical transport properties are observed. Below the room temperature, the resistivity initially shows the behavior of a diffusion-controlling metal-like behavior, corresponding to nearly an unchanged carrier concentration and increased carrier mobility with temperature as shown in Fig. 2. The electrical resistivity starts to deviate from such behavior when temperature goes down to near 160 K and forms a "hump" in temperature range from 160 K down to 80 K. A similar abnormity in *dc* electrical resistivity was characterized as the appearance of CDW fluctuation in $NbSe_3$, $TaS_3$ and $K_{0.3}MoO_3$ [4-9]. Additionally, a hysteresis loop in the temperature dependent electrical resistivity was observed in the range of 80 – 160 K. Conventionally, the sharp peak in the curve of logarithmic derivative of electrical resistivity (ln $\rho$, or log $\rho$) with respect to reciprocal temperature $1/T$, i.e., $d\{\ln \rho(T)\}/d(1/T)$ versus $T$, was used to define the character temperature of the Peierls transition. By applying the same method, we determined the transition temperature $T_c$ is 125 K, as shown in the Fig 1(d). Due to the polycrystalline structure, the Peierls transition process has a widely expanded temperature range from 160 K to 90 K.

According to the mean field theory [2], the elevated electrical resistivity with decreasing temperature, follows the activation energy relationship, i.e., $\rho \propto \exp(\Delta/k_B T)$ where the $\Delta$ is the effective energy gap due to the CDW. The energy gap could be obtained by extracting the slope of the curve of ln $\rho$ vs. $1/T$. Fig. 1(e) shows the numerical calculation of the $\Delta$ by using a differential step value $\delta T = 1$ K, and shows a temperature dependent energy gap of $\Delta = 40.9-0.265T$ (meV). We note that the exact physical meaning of the temperature dependent $\Delta(T)$ is not clear. It could be an average result of all the grains or domains with and without an open gap. The saturated energy gap at zero temperature is estimated to be 40.9 meV by applying the $T \rightarrow 0$ according to the linear relationship. Fig. 1(e) compares the saturated energy gap of Cu$_2$Se with other reported CDW materials as a function of their Peierls transition temperature. The value of $\Delta/k_B T_c$ for all the samples is larger than the theoretical Bardeen-Cooper-Schrieffer (BCS) relationship, i.e., $2\Delta = 3.52 k_B T_{CDW}^{MF}$, which means that the real observed Peierls transition temperature is much lower than the theoretical value based on the mean field theory. Furthermore, the electron-phonon coupling constant $\lambda$ and the coherent length of electron-hole pair $\xi_0$ of the CDW ground state could be estimated within the free electron model by using the following relationships,[2]

$$\Delta = 2\varepsilon_F e^{-1/\lambda} \quad , \tag{1}$$

$$\xi_0 = \frac{\hbar v_F}{\pi |\Delta|} , \qquad (2)$$

$$\varepsilon_F = \frac{\hbar^2}{2m_0}(3\pi^2 n)^{2/3} \qquad (3)$$

$$v_F = \left(\frac{2\varepsilon_F}{m_0}\right)^{1/2} , \qquad (4)$$

where $\hbar$ is the reduced Plank constant, $m_0$ is the free electron mass, and $n$ is the carrier concentration at zero temperature. By using the Hall carrier concentration near 90 K, *i.e.*, $n = 1.6 \times 10^{20}$ cm$^{-3}$, to Eq. (1 – 4), we derive the characterized parameters: $\varepsilon_F = 0.11$ eV, $v_F = 1.94 \times 10^7$ cm s$^{-1}$, $\lambda = 0.6$, and $\xi_0 = 1$ nm. Interestingly, the reported CDW materials are separated into two groups in the map of coupling constant versus transition temperature as shown in Fig. 1(f). Cu$_2$Se falls in the group with NbSe$_3$ and TaS$_3$, while other materials such as K$_{0.3}$MoO$_3$, KCP, and (TaSe$_4$)I get into another group. The coherent length of electron-hole pair $\xi_0$ of Cu$_2$Se is one order of magnitude less than K$_{0.3}$MoO$_3$ and (TaSe$_4$)$_2$I, and also 6 times less than NbSe$_3$. The $\xi_0$ of as-fabricated Cu$_2$Se is much smaller than the grain size of 1 μm, which could explain why we are able to observe the CDW in poly-crystalline Cu$_2$Se. We also show the direct measurement of a cold pressed sample from the Cu$_2$Se nanopowder. It is found that non-linear electrical resistivity due to CDW fluctuation is significantly suppressed. A restored CDW ground state was seen by annealing the cold pressed sample, which demonstrates the behavior of the cold pressed sample as the nano grains merged into micro grains. The typical grain size of our cold pressed sample is in the order of 100 nm, measured by JEOL JSM-6340F Field Emission Scanning Electron Microscope.

Fig. 2 shows a greater detail of Hall measurements of Cu$_2$Se, which demonstrates several complicated features near the Peierls transition. The Hall carrier concentration of Cu$_2$Se, as shown in Fig. 2 (a), is almost a constant as the temperature decreases from room temperature to 160 K, and then gradually decreases from $2.1 \times 10^{20}$ cm$^{-3}$ to $1.6 \times 10^{20}$ cm$^{-3}$ as the temperature further goes down from 160

K to around 90 K, and finally reaches a stable value below 90 K. The decreasing carrier concentration from 160 K to 90 K was interpreted as the appearance of the charge density wave, which corresponds to the formation of electron-hole pairs in real space and a decrease of the area of the Fermi surface in reciprocal space. In addition, in Fig. 2 (b), the carrier mobility also shows three stages in the temperature dependent curve, which is consistent with the variation of the carrier concentration. Near room temperature, the temperature dependent carrier mobility shows a typical degenerated semiconductor behavior with acoustic phonon scattering, *i.e.*, $r$ = 0.8-1.2 in the relationship of $\mu \propto T^{-r}$ at this temperature range. The power index shows a continuous decrease to almost zero when temperature goes from 300 K to 160 K, which may due to the onset of the CDW. During the phase transition temperature range from 160 K to 90 K, carrier mobility is nearly flat. After entering the CDW ground state, the carrier mobility starts to rise again in a more gentle way. The electrical resistivity does not follow the relationship of $\rho \propto T^{-2}$ as expected for a normal metal. The temperature dependent carrier concentration and carrier mobility divide the temperature range into three regions according to the electronic controlling mechanism: diffusive region, mixture region, ballistic-like region. Fig. 2(c-g) shows the I-V curve at different temperatures from these three temperature regions. In the diffusive controlling region, the I-V curve at temperature of 200 K or 160 K shows a good linear behavior. Conventionally, a *dc* I-V curve was used to identify the threshold, or the critical electric field, $E_c$ of the CDW [2, 6, 9]. The I-V curve of materials with CDW ground state usually behaves a linear relationship when the current is small, which corresponds to a constant differential resistance *dV/dI*. As the applied electric field exceeds a threshold, the *dV/dI* becomes significantly decreased with the continuous increasing applied electric field and enters a nonlinear I-V regime. However, we did not observe such nonlinear I-V curve at 90 K. One of the reasons may be due to the high electrical conductance of the $Cu_2Se$ sample we measured (~0.02 Ω), so the highest electric field could be applied is limited by the constant current source. Currently, the highest electric field is only $3\times10^{-5}$ V cm$^{-1}$, which is much less than a typical $E_c$ of $8.7 \times 10^{-2}$ V cm$^{-1}$ for Nb$_3$Se [6] and 1.2 V cm$^{-1}$ for (TaSe$_4$)$_2$I [9]. We also made another thinner Cu$_2$Se sample with a 10 times larger electrical resistance (~0.2 Ω) and raise the applied electric field to ~$3 \times 10^{-4}$ V cm$^{-1}$. However, we still did not observe the decrease in differential electrical resistivity with increasing applied electric field. As we continued to measure the I-V curve at lower temperatures, a wave-like I-V curve was observed near 50 K with a period of ~100 μA and amplitude of 2 μV, as shown in Fig. 2(c). A negative differential resistance $dV/dI$ at each wave was identified. To our best knowledge, such regular oscillations in $dc$ I-V curve have not yet been reported in Cu$_2$Se and also other materials.

Fig. 3 shows more I-V curves of Cu$_2$Se near 50 K measured under different conditions. Fig. 3(a) shows that the wave-like I-V curve is repeatable, not a one-time result, in which the second measurement is conducted several days after the first measurement and with a little shift compared with the first one. In the second measurement, shown in Fig. 1S, the I-V curve at four different temperatures of 200 K, 120 K, 50 K, and 30 K were measured. Only the I-V curve near 50 K shows the notable wave-like feature, while the ones measured at 120 K and 200 K only show normal linear relationship. It is noted that the I-V curve at 30 K still shows a weak oscillation as shown in the Fig.1S. In order to confirm our measurement, we prepared another thin sample with resistance of 0.2 Ω as shown in Fig. 3(b). For the thin sample a narrower temperature range was detected. The wave-like I-V curves was observed at the temperatures of 50 K, 48 K, and 30 K, but not at 52 K and 90 K. Furthermore, the period of the fluctuation is increased as the temperature decreases from 50 K to 30 K. It is suggested that the regular fluctuation is highly sensitive to the temperature. On the other hand, the amplitude of the fluctuation of the thin sample (0.2 Ω) is smaller than the thick one (0.02 Ω), *i.e.,* the differential resistance or resistivity drops below zero for 0.02 Ω while it is still positive for 0.2 Ω, which suggested that the phenomena we observed is also sample size related. In order to exclude the possibility that the signal we observed may come from the equipment, we have done similar I-V curve measurement for a standard resistor (1 Ω) from 30 K to 200 K, but we did not find any wave-like, or any noise-like, but only perfect linear I-V curves, *e.g.* at 50 K, as shown in the Fig. 2S.

The effect of the magnetic field and the electrical field on the wave-like I-V curve was further investigated, as shown in the Fig. 3(c-d). Under the magnetic field of 9 T, the period and amplitude of the fluctuation in the I-V curve is nearly unchanged when the current is less than 500 μA. According to the shift shown in Fig. 3 (a), the superposition of the beginning pars of two curves in Fig. 3 (c) is better regarded as a coincidence. It seems the shift cannot be eliminated, the measurements were performed on the third sample (0.03 Ω) as shown in Fig. 3S, and the shifts were still observed. However, the fluctuation period decreases by a little bit when current is higher than 500 μA. It seems that the electrical field may play a more important role to change the electronic transport behaviors in CDW ground state. We have applied electrical field of $1.5 \times 10^{-4}$ V cm$^{-1}$ on the thin sample (0.2 Ω) over 5 hrs. at 50 K, and then re-measured the I-V curve. Surprisingly, the wave-like feature in the I-V curve disappeared, as shown in Fig. 3(d).

It is well-known that a noise-like fluctuation in the *dV/dI* was widely observed in the NbSe$_3$, TaS$_3$, and K$_{0.3}$MoO$_3$ as a result of the collective moving of the CDW when the applied electric field is higher than the critical field $E_c$ [5, 7, 8]. However, the applied electrical field in our measurement is far less than the threshold value $E_c$. The fluctuation we observed is very regular, rather than an electronic noise. Another noticeable phenomenon is the Shapiro steps in the *dc* I-V curves when an *rf*-frequency *ac* signal was applied to the samples [18-21]. However, these steps can only be observed when the *ac* signal is nonzero due to its interference nature. The Shapiro steps do create a fluctuation in the differential resistance *dV/dI*, but not a negative *dV/dI*. Furthermore, the amplitude of the sub-harmonics decreases with an increasing *dc* electric field. In contrast, our measurement is only conducted under stable *dc* electric field step by step with a step current of 5 μA. At each step, a waiting time of 15 sec was set to wait for the voltage value becoming stable in our constant current measurement mode. The amplitude and period of the wave we observed did not show a notable decay with increasing current. Furthermore, a negative differential resistance was observed at each wave peak. One of the possible mechanisms for the wave-like I-V curve could be related to the special periodic modulation of the electronic charge density in CDW ground state, and also the formation of electron-hole pairs due to the nesting on the Fermi surface, which may result in a local carrier concentration, or carrier mobility jump as the Fermi energy across the gap with rising applied *dc* electric field. We note that the negative differential resistivity phenomenon is usually observed in a molecular conductor, in which a ballistic mechanism is dominant in the electronic transport [23]. Recently, a negative differential resistivity was also reported in a CDW material $BaIrO_3$ single crystal at 4.2 K [24]. However, only one peak was observed in Nakano's I-V curve. The observed negative differential resistivity in $Cu_2Se$, which is related to a regular applied electric field, is still unique. We believed that the onset of wave-like fluctuation of I-V curve in $Cu_2Se$ should be a new material-related phenomenon.

Fig. 4(a) shows the doping effect on CDW fluctuation of $Cu_2Se$. The first investigation was to add extra Se (which is equal to the Cu deficiency) to increase the hole concentration from $2.1 \times 10^{20}$ cm$^{-3}$ to $2.5 \times 10^{20}$ cm$^{-3}$. A similar "hump" in the electrical resistivity was seen throughout the Peierls transition but the temperature was increased from 125 to 138 K. We also note that in an early work on $Cu_{1.7-1.8}Se$ ($p=3.0 \times 10^{21}$ cm$^{-3}$) the data shows a similar but irreversible "hump" in electrical resistivity near 180 K [18]. It seems the increased carrier concentration could influence the transition temperature. Recently, a theoretical study suggested that a phase transition due to the CDW may exist in $Cu_2Se$ near 120 K [18]. Zn has one more valance electron than Cu, while Ni has one valance electron less than Cu, both of which will significantly change the carrier concentration of the $Cu_2Se$. Both samples with Zn and Ni have shown notable Peierls transition near 130 K, which is slightly higher than the pure $Cu_2Se$, and resulted in a carrier concentration and mobility decrease near the Peierls transition as shown in the Fig. 4(b). Furthermore, $Cu_{1.98}Zn_{0.02}Se$ has the largest electron-phonon coupling constant of 0.65, while $Cu_{1.9}Ni_{0.1}Se$ has the largest saturated energy gap of 61.2 meV. The way of Zn and Ni influencing the CDW is different from the Cu vacancy. Besides tuning the electrons, we also partially substituted the Se with Te allowing for increased phonon scattering and also breaking the perfect nesting on Fermi surface. No evident Peierls transition was seen in the sample $Cu_2Se_{0.9}Te_{0.1}$. The suppression of propagation phonon owing to the alloying effect breaks the electron-phonon coupling, meanwhile the substitution of Se by Te in the lattice structure results in the removal of the induced subsequent lattice instability. Fig 4(c) shows the I-V curve of the sample $Cu_{1.98}Zn_{0.02}Se$ is conducted over a wide temperature range, and a similar wave-like fluctuation was still observed at the temperature below 40 K. It seems that the onset temperature of wave-like fluctuation is sensitive to the dopants. The derived wave-like I-V curve of $Cu_{1.98}Zn_{0.02}Se$ at 40K, we clearly see a negative differential resistance, as shown in the Fig. 4 (d). We still have no idea about the clear picture of this negative differential resistance. However, we believed what we observed is a materials related new phenomena, which could be a new quantum effect or strong electron-phonon coupling effect in the CDW ground state. Such kind of wave-like fluctuation in the I-V curve may provide a new way to probe the new electronic transport phenomena of the CDW ground state.

**Conclusions**

A Peierls transition was identified in polycrystalline $Cu_2Se$ near 125 K according to the nonlinear electrical resistivity, carrier concentration and also carrier mobility. The characteristic parameters of the charge density wave in the polycrystalline $Cu_2Se$ was calculated to show a saturated energy gap $\Delta$ of 40.9 meV at zero temperature, the electron-phonon coupling constant $\lambda$ of 0.6 and the coherent length of electron-hole pair of $\xi_0 \sim 1$ nm. After entering the new CDW ground state below 90 K, a negative differential resistivity was identified with a regular fluctuation of the I-V curve. The regular fluctuation in *dc* I-V curve was not magnetic field sensitive, but temperature and sample size sensitive. The wave-like fluctuation is different from the conventional electronic noise observed above the threshold $E_c$, and also the Shaprio step with a finite applied *ac* field. Both the Zn and Ni doped $Cu_2Se$ show CDW characters with increased energy gap and electron-phonon coupling constant. No notable Peierls transition was identified from the temperature dependent resistivity and Hall measurement in Te doped $Cu_2Se$. Similar wave-like I-V curve was also seen in the $Cu_{1.98}Zn_{0.02}Se$ near 40 K, which suggests that the onset of wave-like fluctuation of I-V curve in $Cu_2Se$ should be a material-related new phenomenon.

**Acknowledgements**

This work is supported by "Solid State Solar-Thermal Energy Conversion Center (S3TEC)", an Energy Frontier Research Center funded by the U. S. Department of Energy, Office of Science, Office of Basic Energy Science under award number DE-SC0001299 (G. C. and Z. F. R.). C.O. wishes to thank the Trustees of Boston College.

**References**

[1]. R. E. Thorne, Charge-density-wave sonductor, Physics Today 49, 42 (1996).

[2]. George Grüner, *Density Waves in Solids* (Perseus Publishing, Cambridge, Massachusetts, 2000).

[3]. M. D. Johannes, I. I. Mazin, Fermi Surface nesting and the origin and original of the density waves in metals, Phys. Rev. B 77, 165135 (2008).

[4]. J. Chaussy, P. Haen, J. C. Lasjaunias, P. Monceau, G. Waysand, A. Waintal, A. Meerschaut, P. Molinie, J. Rouxel, Phase transition in $NbSe_3$, Solid State Comm. 20, 759-763 (1976).

[5]. R. M. Fleming, C. C. Grimes, Sliding-mode conductivity in NbSe$_3$: observation of a threshold electric field and conduction noise, Phys. Rev. Lett. 42, 1423 (1979).

[6]. P. Monceau, J. Richard, M. Renard, Charge-density-wave motion in NbSe$_3$. Phys. Rev. B 25, 931-947(1982).

[7]. A. Zettl, G. Gruner, A. H. Thompson, Charge-density-wave transport in orthorhombic TaS$_3$ Phys. Rev. B 26, 5760-5772 (1982)

[8]. M. F. Hundley, A. Zettle, Noise and Shapiro step interference in the charge-density-wave conductor K$_{0.3}$MoO$_3$, Solid State Comm. 66, 253-256 (1988)

[9]. Z. Z. Wang, M. C. Saint-Lager, P. Monceau, M. Renard, P. Greasier, A. Meerschaut, L. Guemas, J. Rouxel, Charge density wave transport in (TaSe$_4$)$_2$I, Solid State Comm. **46**, 325-328 (1983).

[10]. J. R. Neilson, T. M. McQueen, Charge density wave fluctuation, heavy electrons, and superconductivity in KNi$_2$Si$_2$, Phys. Rev. B 87, 045124, (2013).

[11]. D. H. Torchinsky, F. Mahmood, A. T. Bollinger, I. Bozovic, N. G, Nat. Mater. 12, 387-390 (2013)

[12]. R. S. Kwok, G. Grüner & S. E. Brown, Fluctuations and Thermodynamics of the Charge-Density-Wave Phase Transition, *Phys. Rev. Lett.* **65**, 365 (1990).

[13]. D. C. Johnston, M. Maki & G. Grüner, Influence of charge density wave fluctuations on the magnetic susceptibility of the quasi one-dimensional conductor (TaSe$_4$)$_2$I, *Solid State Comm.* **53**, 5 (1985).

[14]. D. Graf, J. S. Brook, E. S. Choi, S. Uji, J. C. Dias, M. Almeida, M. Matos, Suppression of a charge-density-wave ground state in high magnetic fields: Spin and orbital mechanisms, Phys. Rev. B 69, 125113 (2004).

[15]. D. Dominko, et al. J. Appl. Phys. 110, 014947 (2011)

[16]. B. Yu, W. S. Liu, S. Chen, H. Wang, H. Z. Wang, G. Chen, Z. F. Ren, Thermoelectric properties of copper selenide with ordered selenium layer and disordered copper layer, Nano Energy **1**, 3, 472 (2012).

[17]. L. Gulay, M. Daszkiewicz, O. M. Strok & A. Pietraszko, Crystal structure of $Cu_2Se$, *Chem. Met. Alloys* **4**, 200 (2011).

[18]. H. Chi, H. Kim, J. C. Thomas, G. Shi, K.Sun, M. Abeykoon, E. S. Bozin, X. Shi, Q. Li, X. Shi, E.Kioupakis, A. Van der Ven, M. Kaviany, C. Uher, Low-temperature structural and transport anomalies in $Cu_2Se$, *Phys. Rev B*. **89**, 195209 (2014).

[19]. J. Richard, P. Monceau, M. Renard, Chrage-density-wave motion in $NbSe_3$. II dynamical properties, *Phys. Rev B*. **25**, 948 (1982).

[20]. A. Zettl, G. Gruner, Phase coherent in the current-carrying charge-density-wave state: ac-dc coupling experiments in $NbSe_3$, *Phys. Rev B*. **29**, 755 (1984)

[21]. R. E. Thorne, W. G. Lyons, J. W. Lyding, J. R. Tucker and John Bardeen, Charge-density-wave transport in quasi-one-dimensional conductors. I. Current oscillations, *Phys. Rev. B* **35**, 6348 (1987).

[22]. A. A. Sinchenko, P. Monceau, Dynamical transport properties of $NbSe_3$ with simultaneous sliding of both charge-density wave, *Phys. Rev B*. **87**, 045105 (2013)

[23]. N. J. Tao, Electron transport in molecular junctions, Nat. Nanotechnology, 1, 173-181 (2006)

[24]. T. Nakano, I. Terasaki, Giant nonlinear conduction and thyristor-like negative differential resistance in $BaIrO_3$ single crystals, Phys. Rev. B, 73, 195106 (2006).

[25]. T. Ohltani, Y. Tachibana, J. Ogura, T. Miyake, Y. Okada, Y. Yokota, Physical properties and phase transitions of β-Cu Se (0.20<$x$<0.25), J. Alloys Compounds, 279, 136-141 (1998).

**Table 1**, Charge-density-wave related Peierls transition temperature $T_P$, saturated energy gap $E_\Delta$, electron-phonon coupling constant $\lambda$, and the temperature $T_{wave}$ observing the wave-like fluctuation in the I-V curve.

| Sample | $T_P$ [K] | $E_\Delta$ [meV] | $\lambda$ [/] | $T_{wave}$ [K] |
|---|---|---|---|---|
| $Cu_2Se$ | 125 | 40.9 | 0.54 | 50 |
| $Cu_2Se$-Zn | 131 | 54.8 | 0.65 | 40 |
| $Cu_2Se$-Ni | 128 | 61.2 | 0.63 | n.a |
| $Cu_2Se_{1.02}$ | 138 | 40.4 | 0.51 | n.a |

n.a.: not available right now.

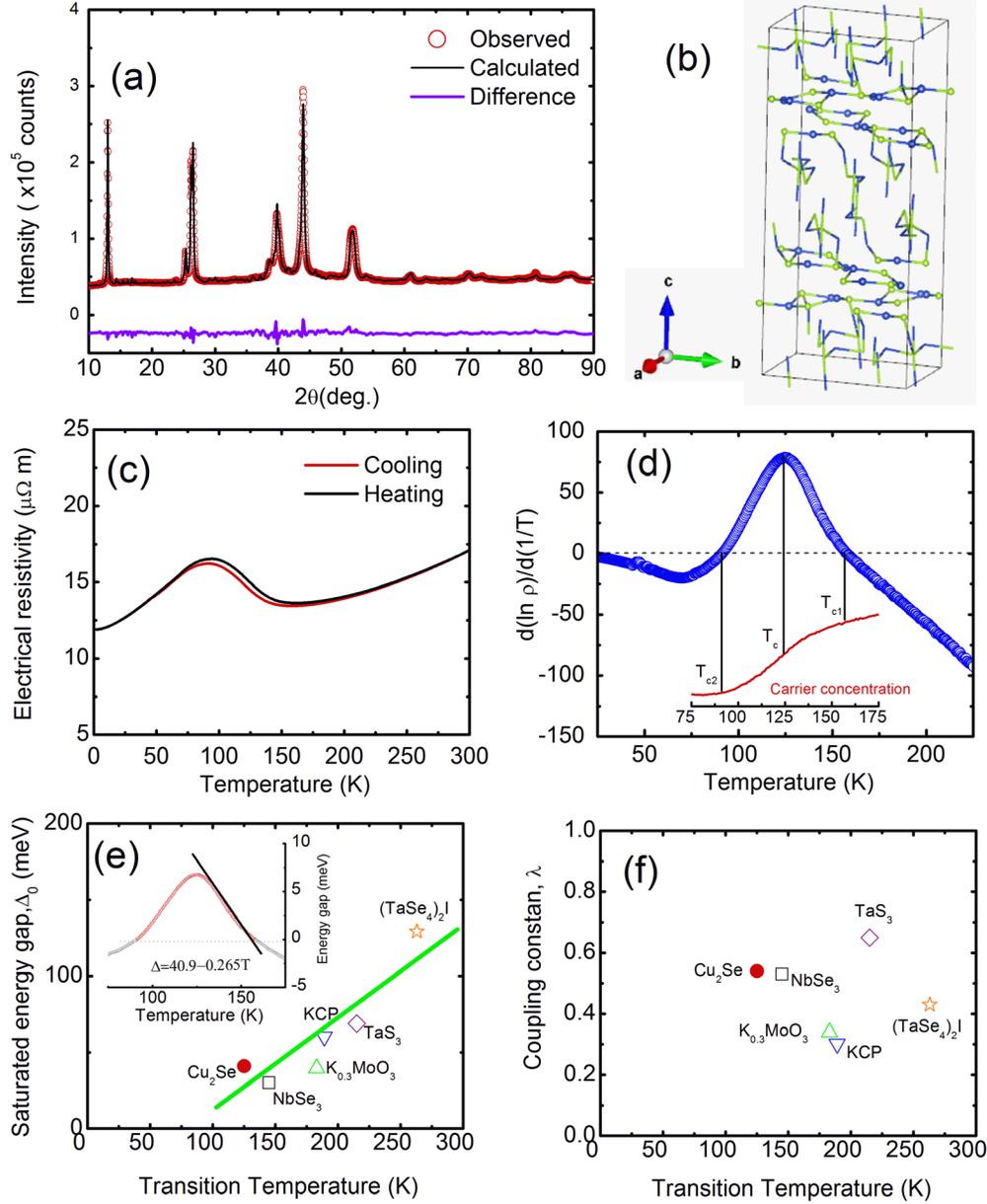

**Figure 1**. (a) Powder XRD pattern of as-fabricated polycrystalline $Cu_2Se$ at room temperature, in which the calculated pattern is based on a monoclinic structure *C2/c* (No. 15). The refine structure ($R_p$ =2.48 %, $R_w$=3.33 %, $R_{exp}$ = 0.44 %) shows the $Cu_2Se$ has a monoclinic unit cell ($a$ = 7.1310 Å, $b$ = 12.3517 Å, $c$ = 27.2880 Å, $α = γ = 90°$, $β = 94.39°$). (b) The atomic structure of monoclinic $Cu_2Se$ in stick mode with hexagonal-ring chain is shown

with the atoms, in which the green ones are Se atoms, blue ones are Cu atoms. (c) Temperature dependent electrical resistivity is measured in both cooling and heating. (d) Temperature dependence of *d(ln ρ)/d(1/T)*. The inset is the varying Hall carrier concentration in the Peierls transition process. (e) Saturated energy gap due to CDW at 0 K as a function of the transition temperature for $Cu_2Se$ and other CDW materials. (f) Electron-phonon coupling constant as a function of the transition temperature for $Cu_2Se$ and other CDW materials. The saturated energy gap and electron-phonon coupling constants of $NbSe_3$, $TaS_3$, $K_{0.03}MoO_3$, KCP, $(TaS_4)_2I$ were adapted from Ref. 2.

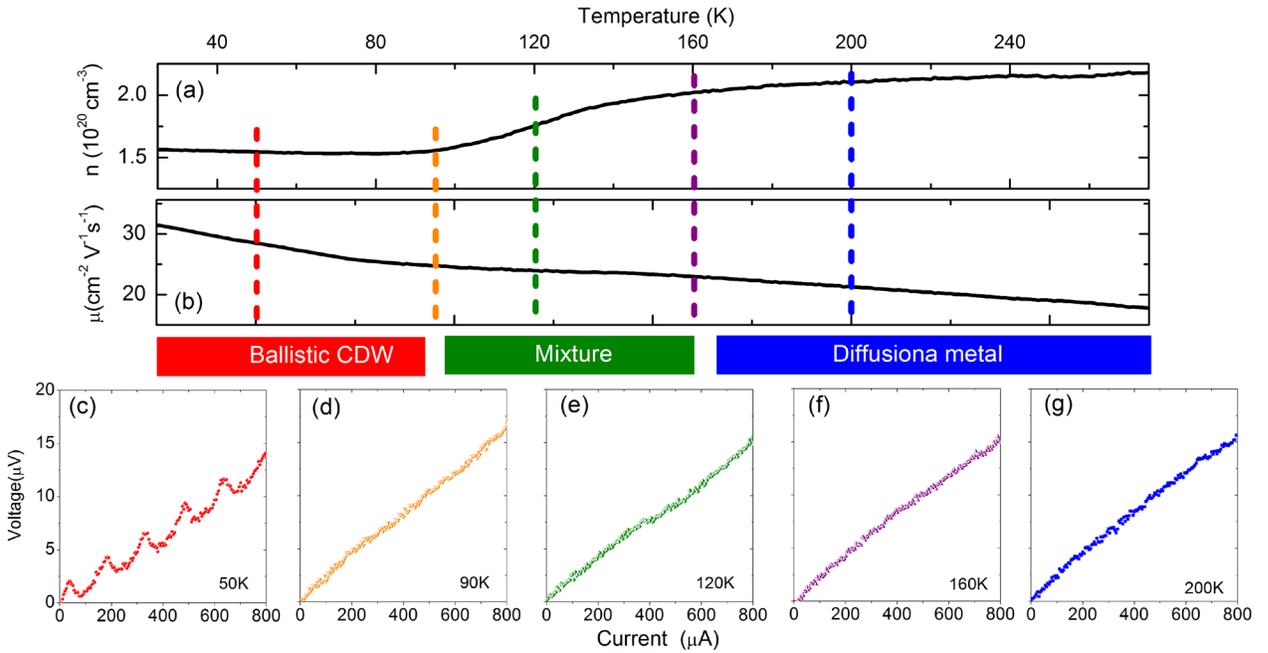

**Figure 2**, Temperature dependent Hall carrier concentration (a) and mobility (b), a standard 4-probe Hall resistance method was performed, measurements were carried out in both +6 T and -6 T magnetic field, in order to eliminate the asymmetric effect from the 2 voltage leads. (c-g) I-V curves at different temperatures, in which regular fluctuations only appear at 50 K, detailed information can be found in the supplementary materials of this article.

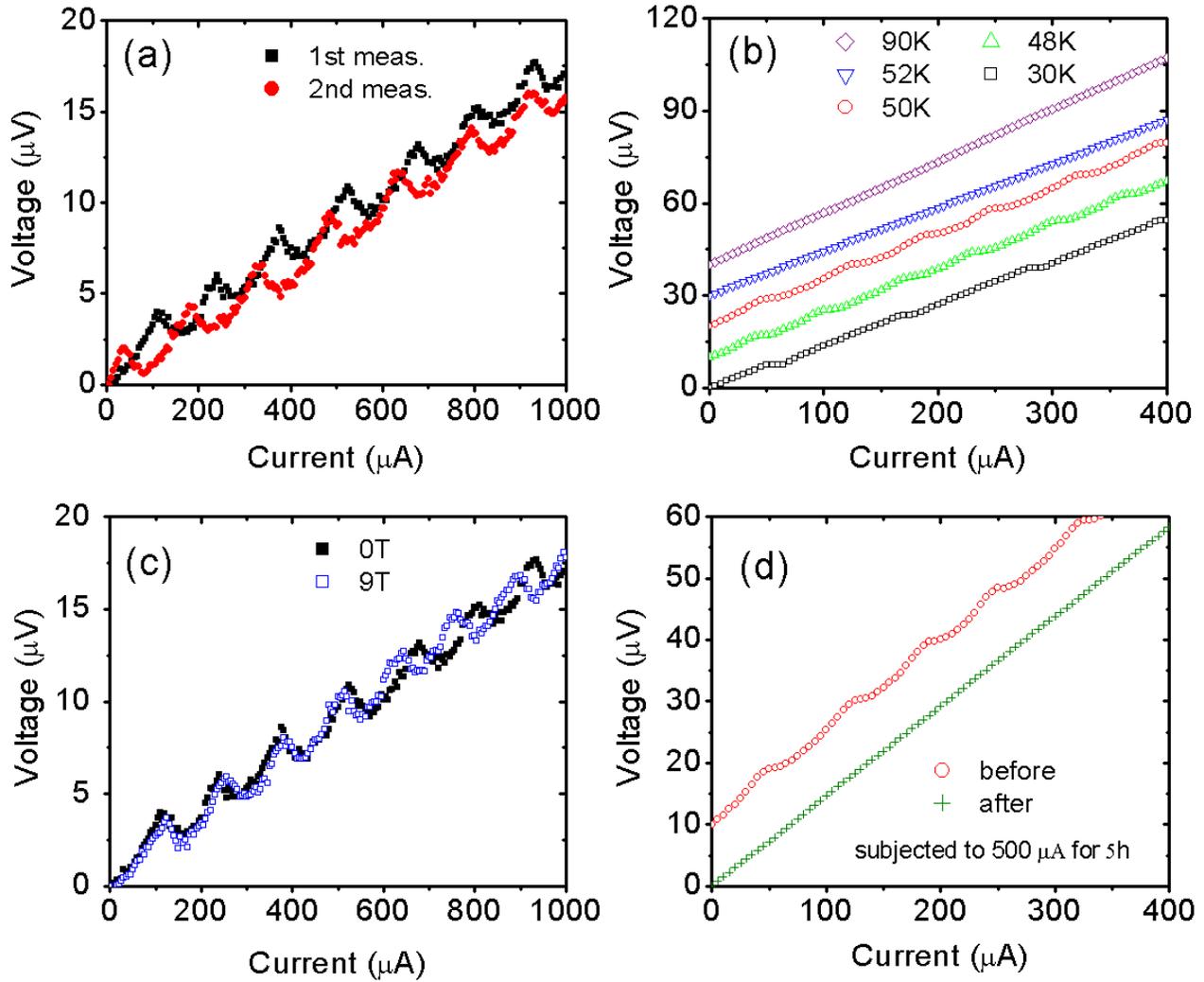

**Figure 3**, Regular fluctuations of I-V curves in the CDW ground state of $Cu_2Se$ polycrystalline. (a) Repeatable wave-like I-V curve of $Cu_2Se$ measured at 50 K for the thick sample with resistance of ~0.02 Ω, (b) I-V curves at different temperatures for the thin $Cu_2Se$ sample with resistance of ~0.2 Ω, which are artificially shifted by 10 μV for 48 K, 20 μV for 50 K, 30 μV for 52 K, and 40 μV for 90 K to separate the I-V curves, respectively, (c) weak magnetic field effect on the I-V curve of the $Cu_2Se$ measured at 50 K for the thick sample with resistance of ~0.02 Ω, (d) effect of the applied large current on the I-V curve of the thin $Cu_2Se$ sample measured at 50 K.

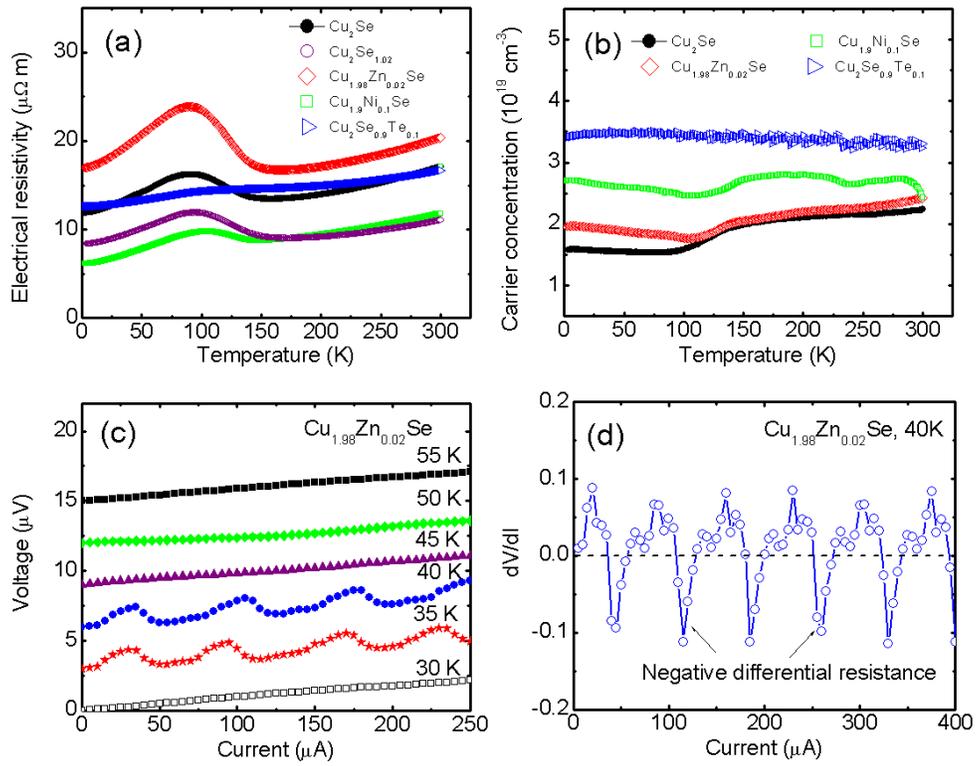

**Figure 4**, Temperature dependent electrical properties of $Cu_2Se$ with different dopants: (a) electrical resistivity, (b) carrier concentration, (c), I-V curves of sample $Cu_{1.98}Zn_{0.02}Se$, (d) differential resistance *dV/dI* of $Cu_{1.98}Zn_{0.02}Se$ at 40 K. The I-V curves in (c) are artificially shifted by 3 μV for 35 K, 6 μV for 40 K, 9 μV for 45 K, and 12 μV for 50 K to separate the I-V curves.

*Supplementary Material for:*

# Anomalous CDW ground state in Cu$_2$Se: a wave-like fluctuation of *dc* I-V curve near 50 K

Mengliang Yao[1‡], Weishu Liu[2‡], Xiang Chen[1], Zhensong Ren[1], Stephen Wilson[1], Zhifeng Ren[2*], Cyril P. Opeil[1*]

[1] Department of Physics, Boston College, Chestnut Hill, MA 02467

[2] Department of Physics and TcSUH, University of Houston, TX 77204

[‡] Equal contributor

[*] Corresponding author: zren@uh.edu, opeil@bc.edu

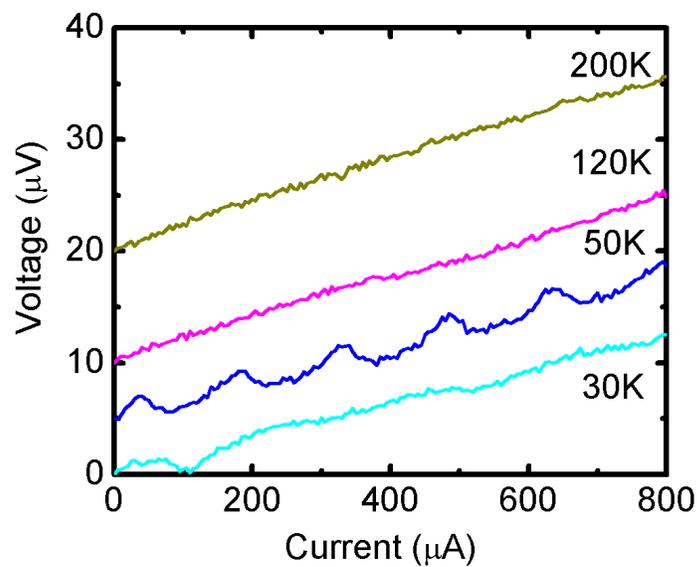

Fig. 1S, Second measurement of the I-V curves of the Cu$_2$Se thick sample (~0.02 Ω) with artificially shifted by 5 μV for 30 K, 10 μV for 50 K, 15 μV for 120 K, and 20 μV for 200 K to separate the I-V curves, respectively.

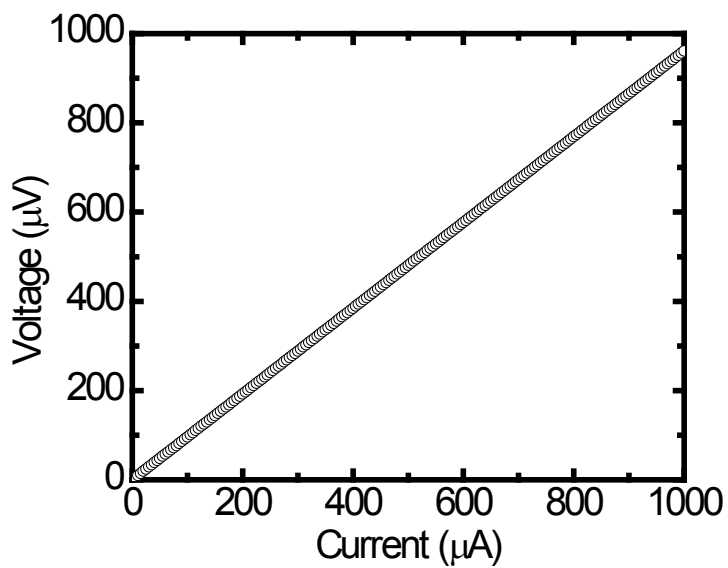

Fig. 2S, The I-V curve of the 1 Ω standard resistor at 50 K, and a straight line was observed. It behaves in a perfect linear relationship, which conclude us that the wave-like fluctuation is the effect from the polycrystalline sample, not from the measuring circuit.

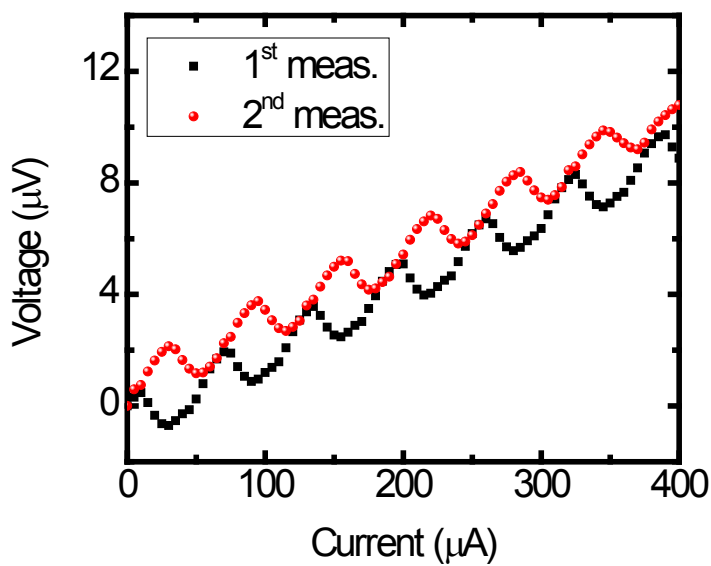

Fig. 3S, Measurements of I-V curves of the third $Cu_2Se$ hot press sample (~0.03 Ω) by Keithley 2182A nano voltmeter. It shows an obvious shift between two measured I-V curves of the third $Cu_2Se$ sample (~0.03 Ω). However, the fluctuation period and amplitude are close.